\documentclass[aps,prd,twocolumn,superscriptaddress,amssymb,eqsecnum,showpacs,showkeyes,secnumarabic,graphics,floatfix,nofootinbib,tightenlines,longbibliography]{revtex4-1}

\usepackage{graphicx}
\usepackage{epsfig}
\usepackage{bm}
\usepackage{cancel}
\usepackage{dcolumn}
\usepackage{amsmath}
\usepackage{hhline}
\usepackage[utf8]{inputenc}
\usepackage[breaklinks=true,colorlinks=true,
linkcolor=blue,urlcolor=blue,citecolor=cyan,
bookmarks=true,bookmarksopenlevel=2]{hyperref}

\def \beq  {\begin{equation}}
\def \eeq  {\end{equation}}
\def \ber  {\begin{eqnarray}}
\def \eer  {\end{eqnarray}}

\begin{document}
\newcommand{\newc}{\newcommand}

\newc{\be}{\begin{equation}}
\newc{\ee}{\end{equation}}
\newc{\ba}{\begin{eqnarray}}
\newc{\ea}{\end{eqnarray}}
\newc{\bea}{\begin{eqnarray*}}
\newc{\eea}{\end{eqnarray*}}
\newc{\D}{\partial}
\newc{\ie}{{\it i.e.} }
\newc{\eg}{{\it e.g.} }
\newc{\etc}{{\it etc.} }
\newc{\etal}{{\it et al.}}
\newc{\lcdm}{$\Lambda$CDM }
\newcommand{\nn}{\nonumber}
\newc{\ra}{\Rightarrow}

\title{The fate of bound systems through Sudden Future Singularities}
\author{L. Perivolaropoulos}\email{leandros@uoi.gr} \footnote{on leave from the Department of Physics, University of Ioannina, 45110 Ioannina, Greece}
\affiliation{Department of Physics, University of Patras, 26500 Patras, Greece}

\date {\today}

\begin{abstract}
Sudden singularities occur in FRW spacetimes when the scale factor remains finite and different from zero while some of its derivatives diverge. After proper rescaling, the scale factor close to such a singularity at $t=0$ takes the form $a(t)=1+ c \vert t \vert^\eta$ (where $c$ and $\eta$ are parameters and $\eta\geq 0$).  We investigate analytically and numerically the geodesics of free and gravitationally bound particles through such sudden singularities.  We find that even though free particle geodesics go through sudden singularities for all $\eta\geq 0$, bound systems get dissociated (destroyed) for a wide range of the parameter $c$. For $\eta <1$ bound particles receive a diverging impulse at the singularity and get dissociated for all positive values of the parameter $c$. For $\eta > 1$ (Sudden Future Singularities (SFS)) bound systems get a finite impulse that depends on the value of $c$ and get dissociated for values of $c$ larger than a critical value $c_{cr}(\eta,\omega_0)>0$ that increases with the value of $\eta$ and the rescaled angular velocity $\omega_0$ of the bound system. We obtain an approximate equation for the analytical estimate of $c_{cr}(\eta,\omega_0)$. We also obtain its accurate form by numerical derivation of the bound system orbits through the singularities. Bound system orbits through Big Brake singularities ($c<0$, $1<\eta<2$) are also derived numerically and are found to get disrupted (deformed) at the singularity. However, they remain bound for all values of the parameter $c$ considered.
\end{abstract}
\maketitle

\section{Introduction}
\label{sec:Introduction}
The observed accelerating expansion \citep{Tsujikawa:2010sc-dark-energy-review,Caldwell:2009ix-dark-energy-review,Copeland:2006wr-review-dark-en} of the universe has opened new windows for possible exotic physics on cosmological scales.  The simplest model, \lcdm \cite{Bull:2015stt-lcdm-review}, based on the existence of a cosmological constant remains consistent with most cosmological observations including the cosmic microwave background (CMB) \cite{Ade:2015xua-planck2015}, baryon acoustic oscillations \citep{Aubourg:2014yra-bao-data,Delubac:2014aqe-bao-data1} large scale velocity flows \cite{Watkins:2014zaa-velocity-flows}  Type Ia supernovae \cite{Betoule:2014frx-jla-snia-data}, growth rate of perturbations data \cite{Huterer:2013xky-growth-data1,Nesseris:2015fqa-growth-data2,Basilakos:2012uu-growth-data2,Nesseris:2007pa-growth-data3}, gamma ray burst data \cite{Izzo:2015vya-grb-dark-energy,Wei:2013xx-grb-dark-energy1,Samushia:2009ib-grb-dark-energy1}, $H(z)$ data \cite{Ding:2015vpa-lcdm-tension-hofz-data}, strong and weak lensing data \cite{Baxter:2016ziy-weak-lensing}, HII galaxy data \cite{Chavez:2016epc-hii-galaxies}, fast radio burst data \cite{Yang:2016zbm-fast-radio-burst}, cluster gas mass fraction data \cite{Allen:2007ue-gas-mass-fraction,Morandi:2016cet-gas-mass-fraction2} etc.  
However, some inconsistencies of \lcdm parameter estimates from specific datasets  are beginning to emerge including inconsistent estimates of the Hubble parameter \citep{Bernal:2016gxb-h0-tension2,Roukema:2016wny-h0-tension1,Abdalla:2014cla-bao-anomaly2,Meng:2015loa-h0-tension2,Qing-Guo:2016ykt-lcdm-h0-tension2,DiValentino:2016hlg} in the context of \lcdm from different datasets, estimates of  the amplitude of the (linear) power spectrum on the scale of 
$8 h^{-1} Mpc$  ($\sigma_8$) \cite{Bull:2015stt-lcdm-review} and estimates of the matter density parameter $\Omega_{0m}$ \cite{Gao:2013pfa-om0-tension}.
In addition to these preliminary observational inconsistencies, there are naturalness theoretical arguments that indicate that physics beyond the standard \lcdm model remain a viable possibility\citep{Tsujikawa:2010sc-dark-energy-review,Caldwell:2009ix-dark-energy-review,Copeland:2006wr-review-dark-en}.

A variety of extensions of \lcdm predict the existence (mostly in the future) of a wide range of singularities\cite{Barrow:2004xh-sfs-first,FernandezJambrina:2006hj-clasify-singularities-Tipler-Krolak,Cattoen:2005dx-visser-classify-milestones-en-cond,Dabrowski:2014fha-sfs-review-classif-varyingconst}. These singularities can be either geodesically incomplete (eg \cite{Caldwell:2003vq-bigrip-defined,Nesseris:2004uj-fate-bound-systems-big-rip,Perivolaropoulos:2004yr-crunch1,Lykkas:2015kls-crunch2}) (geodesics do not continue beyond the singularity and the universe ends at the classical level) or geodesically complete\cite{Dabrowski:2014fha-sfs-review-classif-varyingconst} (geodesics continue beyond the singularity and the universe may remain in existence). 

Geodesically incomplete singularities include the Big Rip\cite{Caldwell:2003vq-bigrip-defined,Nesseris:2004uj-fate-bound-systems-big-rip} where the scale factor diverges at a finite future time due to infinite repunsive forces of phantom dark energy, the Little Rip \citep{Frampton:2011rh-little-rip-models} and the PseudoRip \citep{Frampton:2011aa-pseudo-rip-some-bound-dissociate} where the divergence occurs at the infinite future time. They also include the Big Crunch where the scale factor vanishes due to the strong attractive gravity of future evolved dark energy eg  in quintessence models with negative potentials \citep{Felder:2002jk,Giambo:2015tja,Perivolaropoulos:2004yr-crunch1,Lykkas:2015kls-crunch2}. Modified gravity, quantum effects and cosmological models that violate the cosmological principle have been shown to weaken or eliminate both geodesically complete and geodesically incomplete singularities \cite{Bamba:2009uf-avoid-sfs-fR,Bamba:2012ky-sfs-non-local-gravity+R^2-avoided,Bamba:2012vg-sfs-infT-torsion-gravity-avoided-with-T^b-like-fR,Barrow:2011ub-quantum-part-prod-doesnot-avoid-sfs,Bouhmadi-Lopez:2014jfa-sfs-smoothed-in=Born-Infeld-cosmology-bound-systems-survive-sfs,BouhmadiLopez:2009jk-brane-bigrip-goesto-sfs,BouhmadiLopez:2009pu-sfs-quantum-avoid,Kamenshchik:2012ij-only-strong-sing-bigbangetc-avoided-byquantum-sfs-noquantumavoidance,Dabrowski:2006dd-quantum-cosmology-big-rip-bigbang--avoided,Dabrowski:2013sea-sfs-regularized-by-varying-constants,FernandezJambrina:2008dt-sfs-mod-grav-Fried-Tipler-Krolak,Kamenshchik:2007zj-sfs-avoided-by-quantumeffects-nobigbrake,Nojiri:2008fk-fR-can-avoid-sfs-evenmore-withquantum,Nojiri:2009pf-cure-bigrip-fsf-modgrav-dark-matter-coupling,Sami:2006wj-avoid-sing-loop-quant-cosm-tsujikawa,Singh:2010qa-curved-loop-quantum-cosmo-avoids-many-sing-big-rip-etc}

Geodesically complete singularities involve a divergence of a derivative of the scale factor $a$ while the scale factor remains finite and different from zero. Such singularities may involve divergence of the Ricci scalar ($R=\frac{6}{a^2} \left( {\ddot a}a + {\dot a}^2+k\right)$ for FRW metric) and Riemann tensor components. Despite of this divergence the geodesics are well defined through the time of the singularity and the Tipler and Krolak integrals\citep{Tipler:1977zza,Krolac1986,FernandezJambrina:2006hj-clasify-singularities-Tipler-Krolak} of the Riemann tensor components along the geodesics remain finite in most cases. The Tipler\cite{Tipler:1977zza} integral is defined as
\be
\int_0^\tau d \tau^\prime \int_0^{\tau^\prime } d\tau'' \vert R^i_{0j0}(\tau'') \vert
\label{tiplerint}
\ee
while  the Krolak integral\cite{Krolac1986} is defined as
\be
\int_0^\tau d\tau'  \vert R^i_{0j0}(\tau') \vert
\label{krolacint}
\ee
where $\tau$ is the affine parameter along the geodesic. The components of the Riemann tensor are expressed in a frame that is parallel transported along the geodesics. These integrals express the time integrals of the tidal forces along geodesics. In a cosmological setup a diverging Tipler integral corresponds to a geodesically incomplete singularity (eg Big Rip) while this is not necessarily true for a diverging Krolak integral. 

A finite Krolak integral means that a cosmological comoving observer on a bound system will experience a finite impulse at the singularity and thus it is possible that the bound system will survive through the singularity. On the other hand a diverging Krolak integral implies an infinite impulse which will dissociate all bound systems at the time of the singularity. However, free particle geodesics may go through such singularity. 

Since the Riemann tensor components involve up to second order derivatives of the scale factor, both integrals (\ref{tiplerint}) and (\ref{krolacint}) are finite if the scale factor has finite first derivative at the singularity even if the second derivative diverges.  If however, the first derivative of the scale factor diverges then only the Tippler integral is finite while the Krolak integral diverges at the geodesically complete singularity and bound systems are expected to dissociate due to the infinite impulse they receive at the singularity.  Singularities where the above integrals diverge are strong singularities. 

By solving the Friedman equations with respect to the pressure and density we may translate the possible divergence of the derivatives of the scale factor at the geodesically complete singularities to divergence of the density and pressure as well as to possible violation of energy conditions. Thus using the equations
\ba
\rho(t)&=&\frac{3}{8\pi G} \left( \frac{\dot a^2}{a^2}+ \frac{k}{a^2}\right) \\
p(t)&=&\frac{1}{8\pi G} \left( 2\frac{\ddot a}{a} +\frac{\dot a^2}{a^2}+ \frac{k}{a^2}\right)
\ea
it becomes clear that when the first derivative of the scale factor is finite at the singularity but the second derivative diverges (Sudden Future Singularities (SFS) \cite{Barrow:2004xh-sfs-first}) the density is finite but the pressure diverges. 

Near a geodesically complete singularity occurring (with no loss of generality) at coordinate time $t=0$, the scale factor after proper rescaling may be expressed in the form\citep{Cattoen:2005dx-visser-classify-milestones-en-cond,FernandezJambrina:2007sx-sfs-proper-time-issues}
\be
a(t)=1+ c \vert t \vert^\eta
\label{scfactans}
\ee
where $c$ and $\eta$ are parameters and for geodesic completeness we assume $\eta\geq 0$. For $0<\eta <1$ the first derivative (and higher) of the scale factor diverges at the singularity (finite scale factor singularity) while for $1<\eta <2$
the second derivative (and higher) diverge at the singularity (SFS). For $c<0$ and $1<\eta <2$ the SFS is known as Big Brake \cite{Kamenshchik:2012ij-only-strong-sing-bigbangetc-avoided-byquantum-sfs-noquantumavoidance,Kamenshchik:2007zj-sfs-avoided-by-quantumeffects-nobigbrake} due to the negative sign of the diverging second derivative (deceleration) of the scale factor.

Comoving free particle geodesics in a FRW metric approaching a geodesically complete singularity are easily obtained by solving the geodesic equation for the radial coordinate which may be written as 
\be
{\ddot r}=\frac{\ddot a}{a} r=\frac{c\;\eta(\eta-1)\;\vert t\vert^{\eta-2}}{(c\;\vert t\vert^{\eta}+1)}r
\label{freegeod}
\ee
where we used eq. (\ref{scfactans}). Eq. (\ref{freegeod}) may also be trivially obtained by demanding that the $\ddot \rho =0$ where $\rho \equiv \frac{r}{a}$ is the comoving coordinate of a comoving observer (not to be confused with the density).
As will be discussed in the next section, eq. (\ref{freegeod}) has finite well behaved solutions for all $\eta \geq 0$ (finite scale factor at the singularity $t=0$) even though the expansion `force' $\frac{\ddot a}{a} r$ and the first derivative of the scale factor may diverge at the singularity. Therefore all singularities involving a finite scale factor are geodesically complete \citep{FernandezJambrina:2004yy-geodesics-sfs-smooth}.

Geodesically complete singularities where the scale factor behaves like eq. (\ref{scfactans}) are obtained in various physical models including quintessence pontentials of the form \cite{Barrow:2015sga-phys-model} 
\be
V(\phi) = A \phi^n
\label{physmod1}
\ee 
with $0<n<1$ and $A$ a constant. In this class of models, it may be shown that when $\phi=0$ the first and second derivatives of the scale factor are finite while the third derivative diverges. This behaviour corresponds to $2<\eta<3$ in eq. (\ref{scfactans}). Other physical models with geodesically complete singularities include tachyonic models \cite{Keresztes:2010fi-tacyonic-bigbrake-tobogcrunch-obs-constr}, modified gravity \cite{Nojiri:2009pf-cure-bigrip-fsf-modgrav-dark-matter-coupling}, loop quantum gravity \cite{Singh:2010qa-curved-loop-quantum-cosmo-avoids-many-sing-big-rip-etc}, anti-Chaplygin gas  \citep{Keresztes:2012zn-sfs-crossing-geodesics-with-delta-function-pressure}, brane models \cite{BouhmadiLopez:2009jk-brane-bigrip-goesto-sfs} etc.

The presence of geodesically complete singularities in our past light-cone is in principle possible and consistent with current observational data. Constraints on such abrupt events have been obtained in Refs. \cite{Park:2015qya-dark-energy-spike,DeFelice:2012vd-rapid-transitions} using standard ruler and standard candle cosmological data constraining the form of the past expansion history of the universe. The possible existence of such events in the future light cone has also been investigated under specific assumptions of the functional form of the future Hubble expansion rate \cite{Lazkoz:2016hmh-sfs-obs-constr-occurs-soon,Yurov:2007tw-obs-constr-bigfreeze-adotinfty-chaplyginalone-insufficient,Jimenez:2016sgs-obs-cons-approach-sfs,Ghodsi:2011wu-sfs-obs-cons-cmb-probs,Denkiewicz:2012bz-sfs-obs-constr-may-occur-near-fut,Denkiewicz:2015nai-sfs-data-growth-test,Dabrowski:2007ci-obs-cons-sfs}.

An important effect of geodesically complete singularities is the disruption or dissociation of bound systems. Geodesically complete singularities with diverging second time derivative but finite first derivative (SFS corresponding to $1<\eta<2$) of the scale factor induce a finite impulse on geodesics which disrupts and may even dissociate bound systems for large enough impulse (large values of $c$ in eq. (\ref{scfactans})). In cases where the first derivative is diverging the induced impulse is infinite and all bound systems dissociate. 

Signatures of bound system disruption due to SFS may be observable in galaxies or clusters leading to additional constraints on the possible existence of such abrupt events in our past light cone. The goal of the present study is to use geodesic equations in order to identify the type of distortion induced on bound systems by SFS. We will also identify the range of parameters for which the distortion of the bound systems is large enough to lead to dissociation. 

The structure of this paper is the following: In the next section we review the derivation of the gravitationally bound particle geodesics in an expanding background and in physical coordinates. The properties of these equations at the SFS is also reviewed and the special case of a free particle is identified. In section III, the free particle geodesics are obtained by solving the geodesic equation both analytically and numerically for specific initial conditions. The geodesics corresponding to a bound particle going through a SFS are obtained numerically in section IV as a function of the parameters $c$, $\eta$ and the angular velocity $\omega_0$ of the bound particle. The range of parameters that lead to dissociation of the bound systems is identified and the form of the geodesics for both dissociated and disrupted systems is obtained. Finally in section V we summarise and discuss possible extensions of the present analysis.

\section{Geodesic equations in physical coordinates}
\label{sec:Section 2}

The metric describing the spacetime around a point mass $M$ embedded in an expanding background in the Newtonian limit (weak field, low velocities) is of the form \citep{Nesseris:2004uj-fate-bound-systems-big-rip}
\be ds^2=(1-\frac{2GM}{a(t)\rho})\cdot dt^2-a(t)^2\cdot
(d\rho^2+\rho^2\cdot (d\theta^2+sin^2\theta d\varphi^2))
\label{met} \ee
This metric is adequate for our analysis as long as $\frac{2GM}{a(t)\rho}\ll 1$ which is consistent with our assumption of a finite scale factor.
Using physical coordinates
\be r=a(t) \cdot \rho \ee it is straightforward to obtain the  geodesics corresponding to metric (\ref{met}) as \be (\ddot{r}-{\ddot{a}\over
a}r)+{GM \over r^2}-r\dot{\varphi}^2=0 \label{geodr} \ee and \be
r^2\dot{\varphi}=L \label{geodf} \ee where $L$ is the conserved
angular momentum per unit mass. 

Using equations (\ref{geodf}) and (\ref{geodr}) we obtain the radial equation of motion of a test particle as

\be \ddot{r}={\ddot{a}\over a}r + {L^2 \over r^3}-{GM \over r^2}
\label{radeqm1} \ee

We now rescale eq. (\ref{radeqm1}) using a time scale $t_0$  (initial time) and a spatial scale $r_0$ (initial circular orbit radius). We also define ${\dot \varphi}(t_0)= \omega_0\equiv {{GM}\over {r_0^3}}$ (initial angular velocity ignoring expansion). After setting ${\bar r}\equiv {r\over {r_0}}$, ${\bar
{\omega_0}}\equiv \omega_0 t_0$ and ${\bar t}\equiv {t\over t_0}$ eq. (\ref{radeqm1}) becomes \cite{Nesseris:2004uj-fate-bound-systems-big-rip,Faraoni:2007es-bound-systems-expanding}

 \be {\ddot {\bar r}}-{{\bar
\omega_0}^2\over {{\bar r}^3}} + {{\bar {\omega_0}}^2\over {{\bar
r}^2}}-{{\ddot a}\over {a}}{\bar r}=0
\label{dleqm} \ee 
In what follows we will omit the bar 
for convenience but we keep using dimensionless quantities. 

In the spacial case where we have no expansion ($a=1$) the solution of eq. (\ref{dleqm}) is $r=1$ ie a circular orbit with unit radius and dimensionless angular velocity $\omega_0$.

We now assume a scale factor that approaches a geodesically complete singularity. Using eq. (\ref{scfactans}) in (\ref{dleqm}) we find the geodesic equation
\be {\ddot {r}}={{
\omega_0}^2\over {{ r}^3}} - {{ {\omega_0}}^2\over {{
r}^2}}+\frac{c\;\eta(\eta-1)\;\vert t\vert^{\eta-2}}{(c\;\vert t\vert^{\eta}+1)}r
\label{sfsgeod} \ee 

\begin{figure}[!t]
\centering
\vspace{0cm}\rotatebox{0}{\vspace{0cm}\hspace{0cm}\resizebox{0.49\textwidth}{!}{\includegraphics{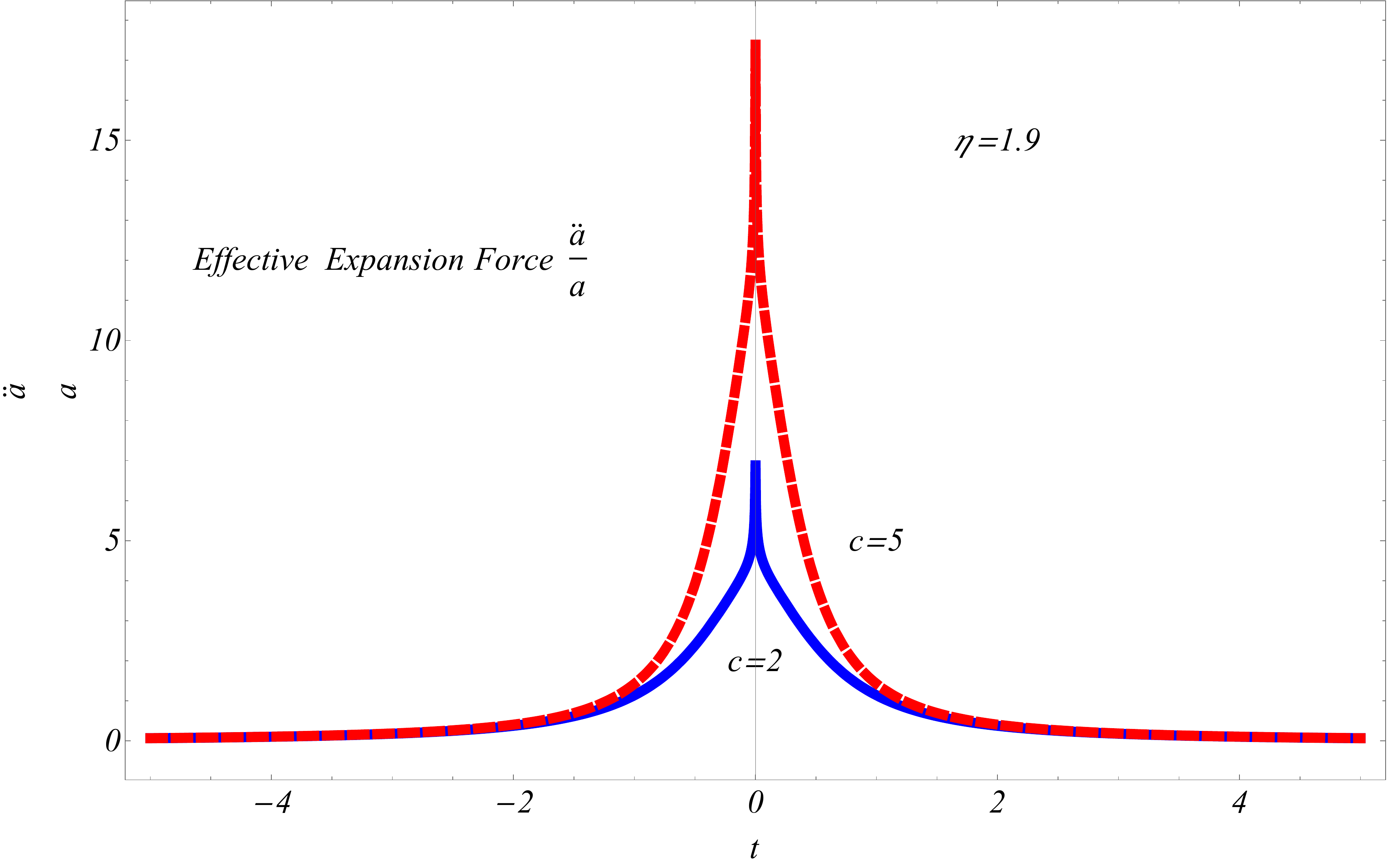}}}
\caption{The effective force $\frac{\ddot{a}}{a}$ due to the expansion for $c=5$ and $c=2$. It diverges at the time of the sudden singularity but its impulse $\int \frac{\ddot{a}}{a} dt$ is finite for $\eta>1$.}
\label{fig1}
\end{figure}

In the special case of a free particle ($\omega_0=0$) the geodesic equation (\ref{sfsgeod}) reduces to (\ref{freegeod}) as expected. This equation is not well defined at the singularity $t=0$ for $\eta<2$ due to the divergence of the expansion force (last term in (\ref{sfsgeod}) shown in Fig. \ref{fig1}. However, when transformed to comoving coordinates $\rho \equiv r/a$ equation (\ref{freegeod}) is written as $\ddot \rho=0$ and the divergence at the singularity disappears in both comoving and physical coordinates given that the scale factor is finite at the singularity. Thus, in comoving coordinates, the geodesic equation is well defined at all times. In addition, the solution is finite on the singularity in both physical and comoving coordinates as discussed in the next section.

\section{Free particle geodesics through sudden singularities}
\label{sec:Section 3}

\begin{figure}[!t]
\centering
\vspace{0cm}\rotatebox{0}{\vspace{0cm}\hspace{0cm}\resizebox{0.49\textwidth}{!}{\includegraphics{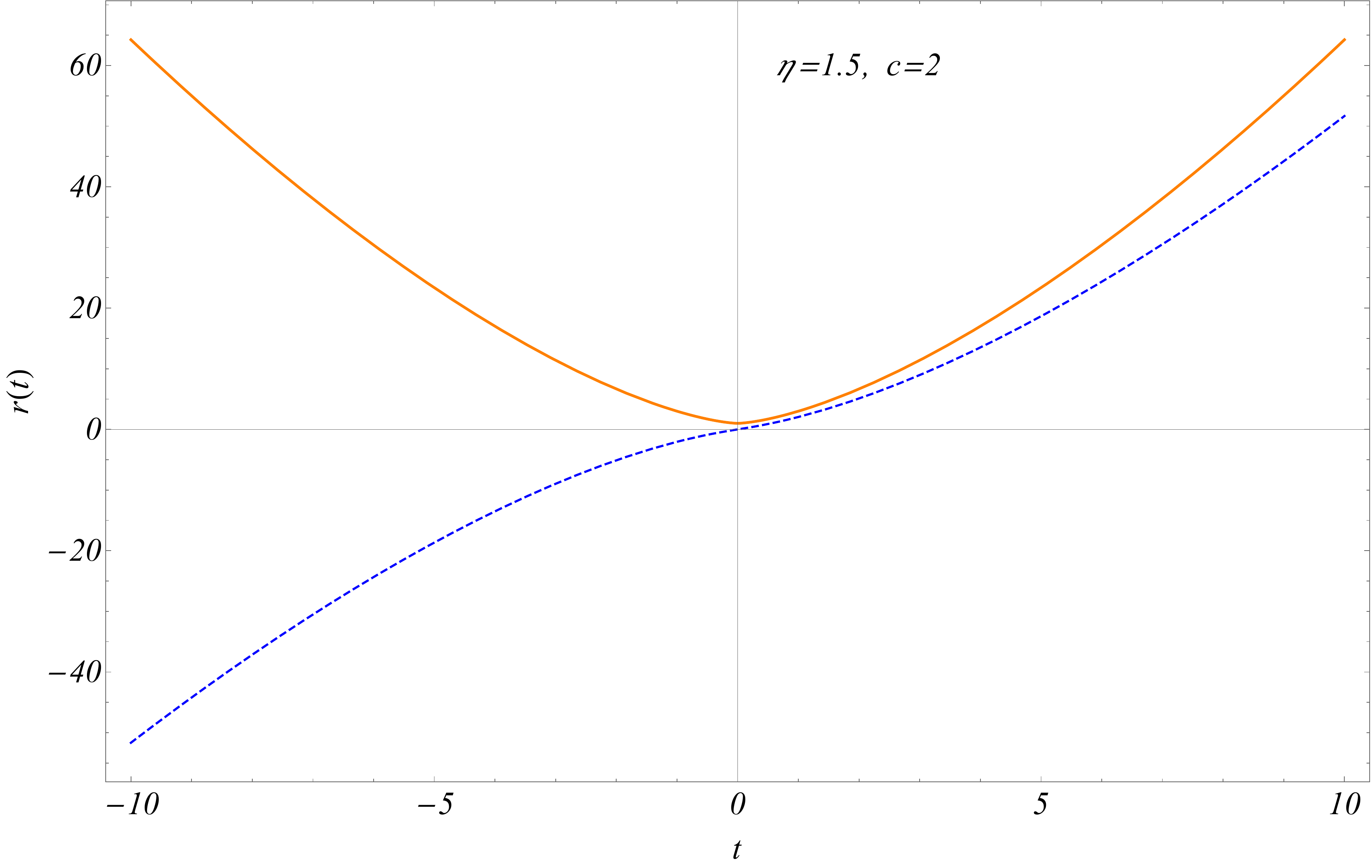}}}
\caption{The two linearly independent solutions for the geodesics of a free particle in a universe with a sudden singularity at $t=0$ ($\eta=1.5$, $c=2$).}
\label{fig2}
\end{figure}

The general solution of the free particle geodesic equation (\ref{freegeod})  through the singularity is a superposition of two independent solutions each with definite parity (one even and one odd). It is of the form
\be
r(t)=A \; (1+c\vert t\vert^\eta) + B\; t \; _2F_1(1,-1+\frac{1}{\eta},1+\frac{1}{\eta},-c\vert t\vert^\eta)
\label{freesol}
\ee
where $A$, $B$ are constants to be determined from the initial conditions and the Hypergeometric function $_2F_1$ is defined as
\be
_2F_1(a,b;c;z)=\sum _{k=0}^{\infty } \frac{a_k b_k z^k}{k! c_k}
\label{hypergeodef} 
\ee
where $a_0=1$, $a_k\equiv a(1+a)(2+a)...(k+a-1)$ and similarly for $b$ and $c$.
The two linearly independent solutions of eq. (\ref{freesol}) are shown in Fig. \ref{fig2} for specific parameter values. They are both finite for all values of $\eta\geq 0$.

\begin{figure}[!t]
\centering
\vspace{0cm}\rotatebox{0}{\vspace{0cm}\hspace{0cm}\resizebox{0.49\textwidth}{!}{\includegraphics{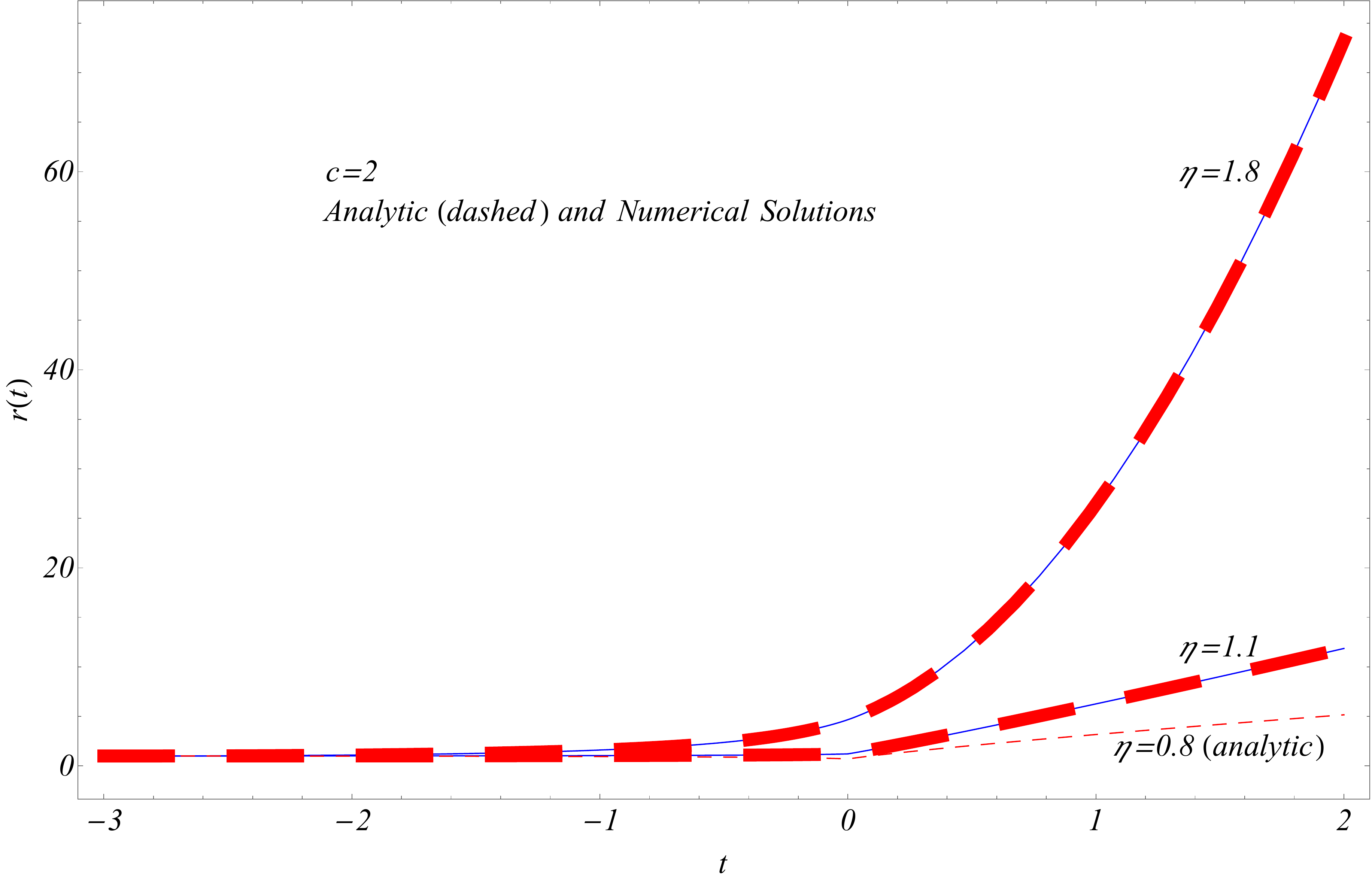}}}
\caption{The geodesic solution $r(t)$ for a free particle with initial conditions $r=1$, ${\dot r}=1$. A sudden singularity background (eq. (\ref{scfactans}))  with $c=2$ and various values of $\eta$ was assumed. The analytic solution (dashed line ) is identical with the numerical solution (thin continuous line). For $\eta<1$ the numerical integration was not possible due to the divergence of the expansion impulse. However, the analytical solution exists even though its time derivative diverges at the singularity (see also Fig. \ref{fig4}).}
\label{fig3}
\end{figure}

Despite of the divergence of the expansion force of eq. (\ref{freegeod}), the numerical solution can be obtained for parameter values where the impulse is finite ($\eta>1$). The numerical solution (thin continous lines) is shown in Fig. \ref{fig3} superposed with the analytical solution (dashed lines) for various values of $\eta$ with initial conditions corresponding to $r=1$, $\dot r=0$. The agreement between numerical and analytical solutions is very good.  For these particular initial conditions the expansion of radius is initially slow (a superposition of the two independent solutions of Fig. \ref{fig2}). However, the impulse at the singularity induces more rapid expansion (in accordance with the evolution of the two independent solutions) which grows faster for larger values of $\eta$ as shown in Fig. \ref{fig3}.

As expected, the geodesic solution remains finite at all times including the singularity at $t=0$ even for values of $\eta$ that correspond to an infinite impulse from the expansion force ($0<\eta<1$). In this parameter range however the divergence of the expansion impulse did not allow the derivation of the numerical solution and thus we have only ploted the analytical solution for $\eta=0.8$ (lowest dashed curve in Fig. \ref{fig3}). 

\begin{figure*}[ht]
\centering
\begin{center}
$\begin{array}{@{\hspace{-0.10in}}c@{\hspace{0.0in}}c}
\multicolumn{1}{l}{\mbox{}} &
\multicolumn{1}{l}{\mbox{}} \\ [-0.2in]
\epsfxsize=3.3in
\epsffile{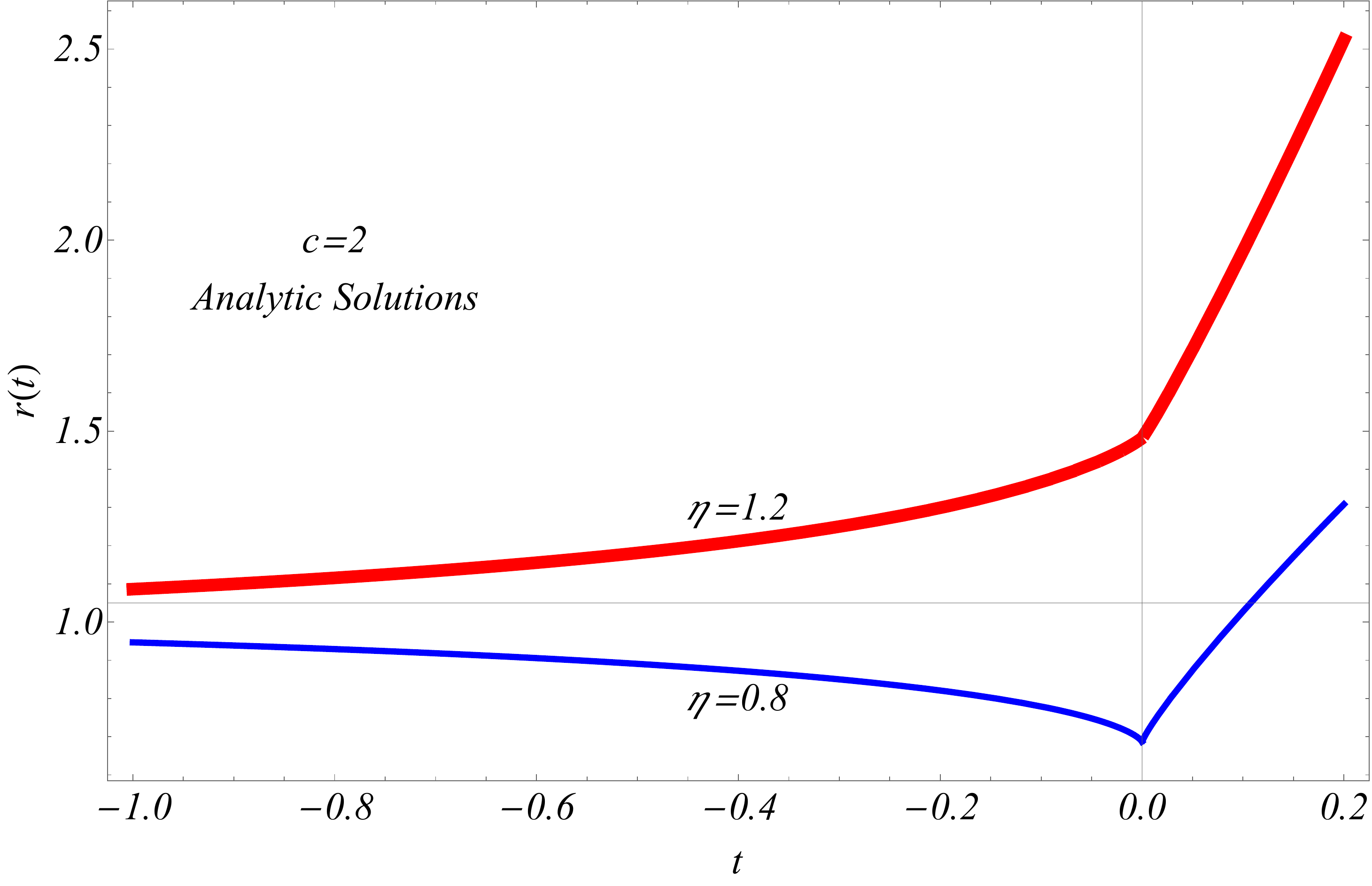} &
\epsfxsize=3.3in
\epsffile{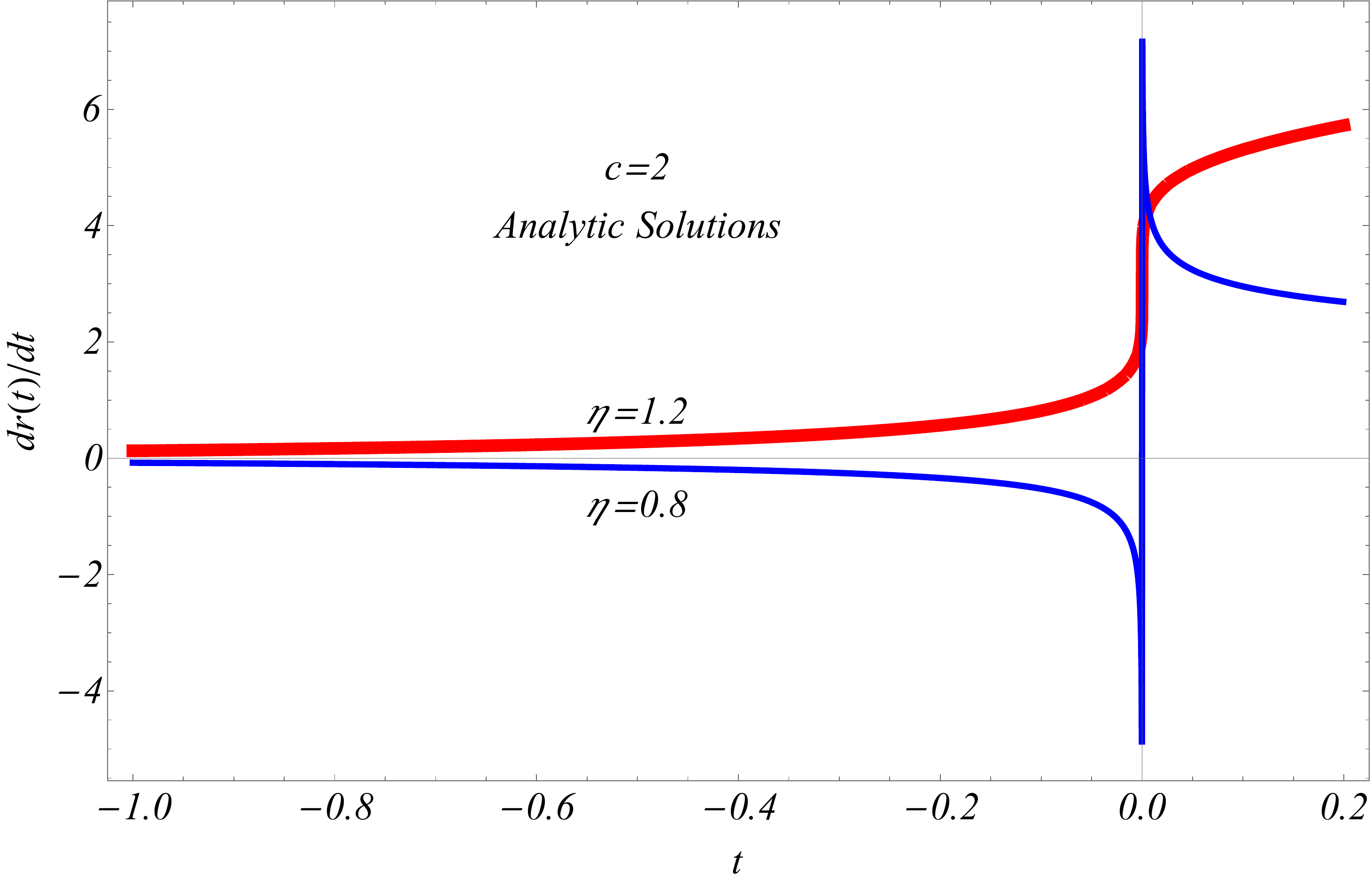} \\
\end{array}$
\end{center}
\vspace{0.0cm}
\caption{\small a: Analytic solution of a free particle geodesic with initial conditions $r=1$, ${\dot r}=0$ at $t_i=-10$ for values of $\eta$ larger and smaller than 1. b: The corresponding time derivative of the geodesic. For $\eta<1$ the magnitude of the discontinuity of the velocity diverges at the singularity.  }
\label{fig4}
\end{figure*}

The effects of the diverging impulse on the analytic solution are shown in Fig. \ref{fig4} where we show in more detail near the singularity, the analytic solution $r(t)$ and its time derivative as a function of time through the singularity. Clearly for $\eta<1$ the force impulse diverges and so does the discontinuity of the velocity $\dot r$ (blue line in Fig. \ref{fig4}b) while for $\eta>1$ the discontinuity remains finite (red line in Fig. \ref{fig4}b). Despite of these discontinuities the analytical solution $r(t)$ is well defined in both cases even though its derivative diverges at the singularity for $\eta<1$. 

In the context of a bound system with the same initial condition, the expansion of the orbit after the singularity induced impulse could be eventually reversed by the gravitational attraction resulting in a deformed bound orbit. This reversal however is not possible for large enough induced impulse and in this case the bound  system would get dissociated. These phenomena will be investigated in detail in the next section.

\section{Bound Systems: Dissociation or Disruption?}
\label{sec:Section 4}

We now proceed to the full solution of the bound system geodesic equation (\ref{sfsgeod}) through the geodesically complete singularity for various parameter values. For $\omega_0\neq 0$ there is no analytical solution to eq. (\ref{sfsgeod}) even though this equation is almost identical to the modified Yermakov's equation\citep{handbookode}. We thus solve the rescaled geodesic equation (\ref{sfsgeod}) with initial condition corresponding to a circular orbit. We set $r(t_i)=s$ and $\dot r(t_i)=0$ where $s$ is a stable equilibrium point at $t_i$ obtained as root of the effective force at the RHS of eq. (\ref{sfsgeod}).
Thus $s$ is the minimum of the effective potential
\be
V_{eff}=-\frac{\omega_0^2}{r}+\frac{\omega_0^2}{2 r^2}-\frac{c\;\eta(\eta-1)\;\vert t\vert^{\eta-2}}{(c\;\vert t\vert^{\eta}+1)} r^2
\label{veff}
\ee
Since the effects of the expansion are initially unimportant in comparison with the gravitational forces, $s$ is close to unity. These initial conditions correspond to an initially circular bound orbit. The impulse of the expansion force shown in Fig. \ref{fig1} is expected to disrupt this circular orbit towards an elliptic orbit or if it can provide enough energy, to dissociate it to a free particle orbit. The angular solution $\varphi(t)$ is obtained from the solution $r(t)$ by integrating eq. (\ref{geodf}) leading to the bound system orbit evolving through the singularity.

\begin{figure*}[ht]
\centering
\begin{center}
$\begin{array}{@{\hspace{-0.10in}}c@{\hspace{0.0in}}c}
\multicolumn{1}{l}{\mbox{}} &
\multicolumn{1}{l}{\mbox{}} \\ [-0.2in]
\epsfxsize=3.3in
\epsffile{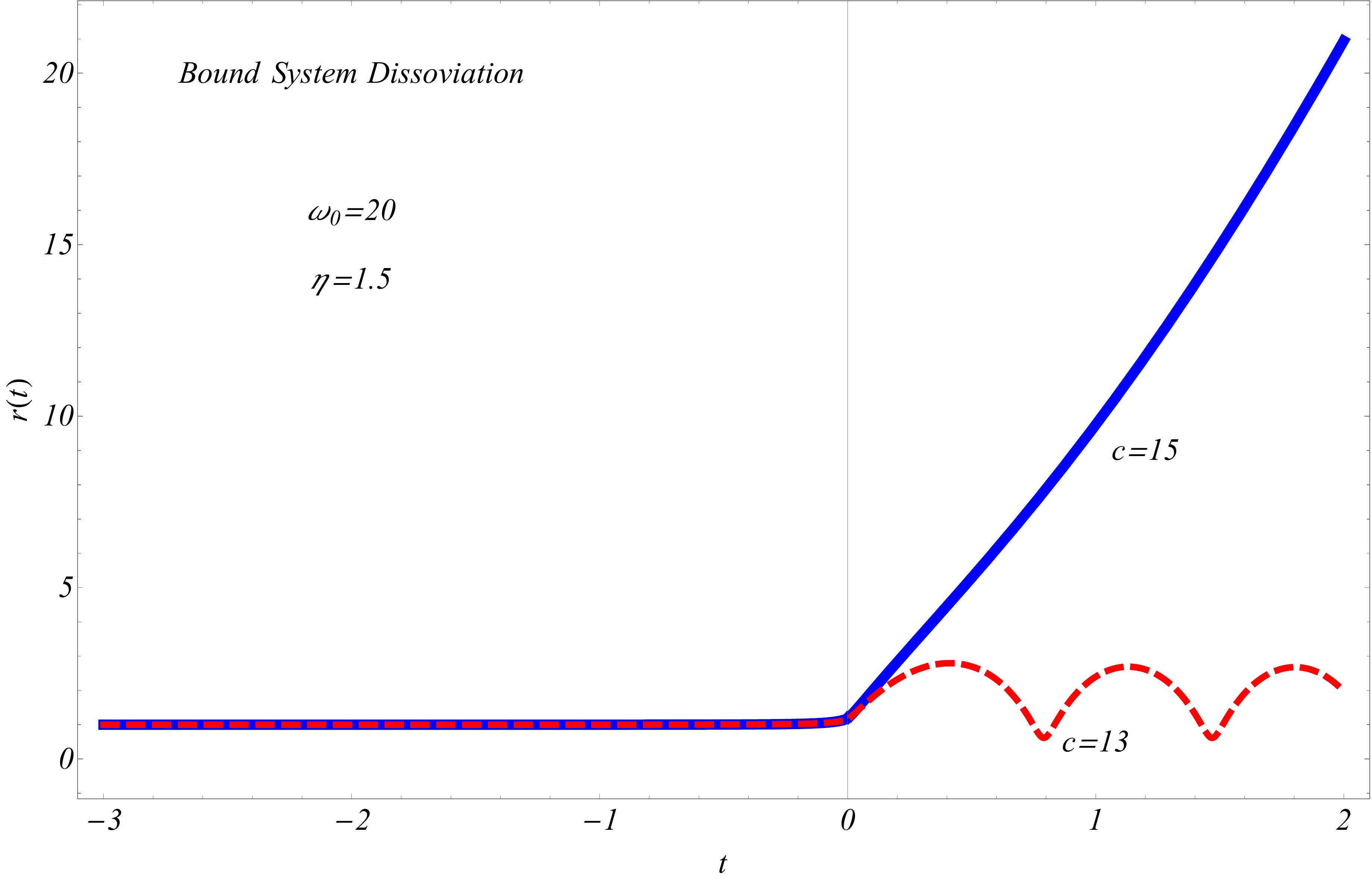} &
\epsfxsize=3.3in
\epsffile{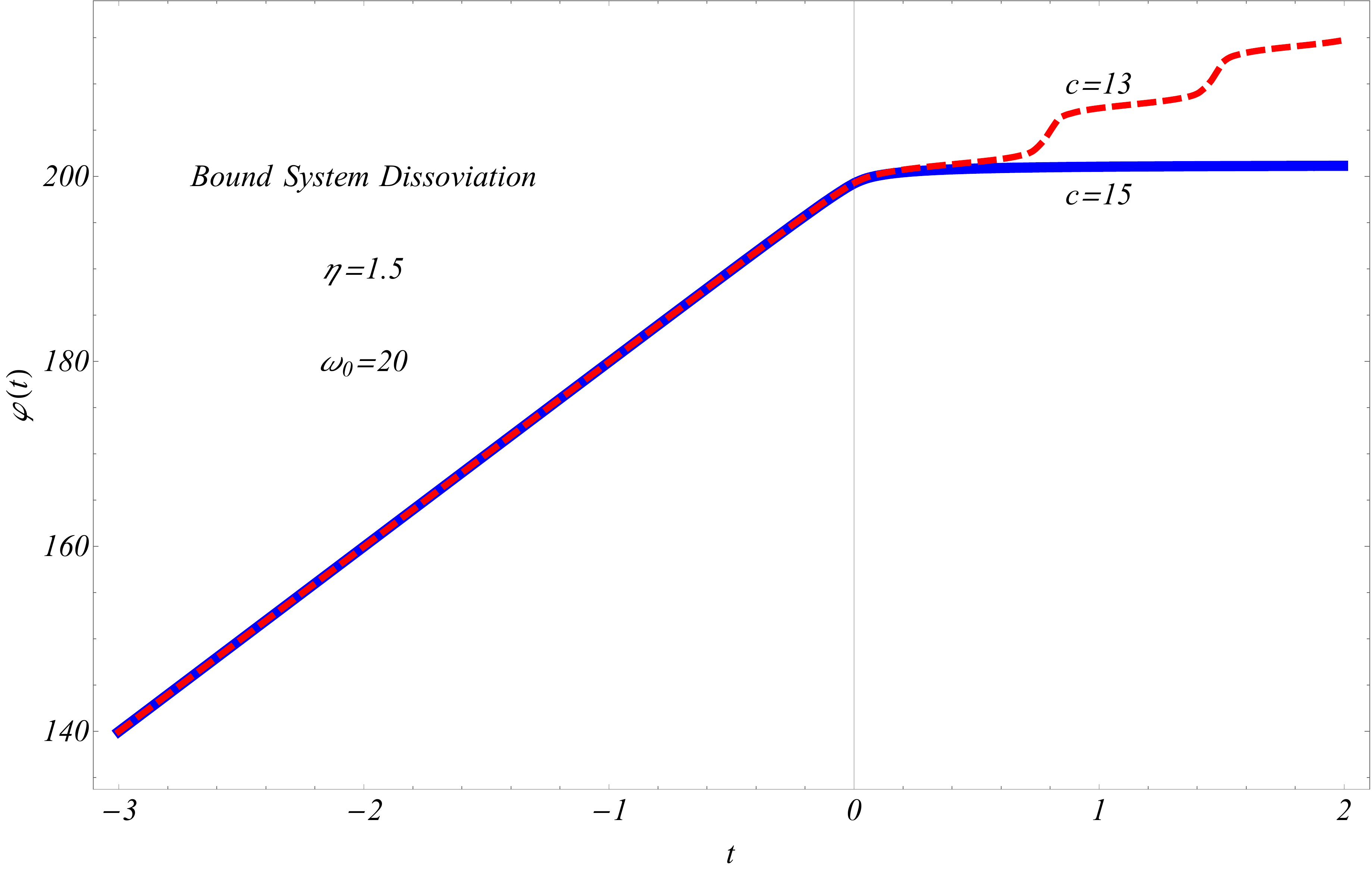} \\
\end{array}$
\end{center}
\vspace{0.0cm}
\caption{\small a: The evolution of the geodesic radial coordinate $r(t)$ for parameter value $c$ above ($c=15$) and below ($c=13$) the critical value $c\simeq 14$ for bound system dissociation. The dissociation of the bound system is seen due to the monotonic increase of $r(t)$ (continous thick blue line).   b: The corresponding plot for the angular geodesic coordinate $\varphi(t)$. When the system dissociates the angular coordinate stops increasing.  }
\label{fig5}
\end{figure*}

\begin{figure}[!t]
\centering
\vspace{0cm}\rotatebox{0}{\vspace{0cm}\hspace{0cm}\resizebox{0.49\textwidth}{!}{\includegraphics{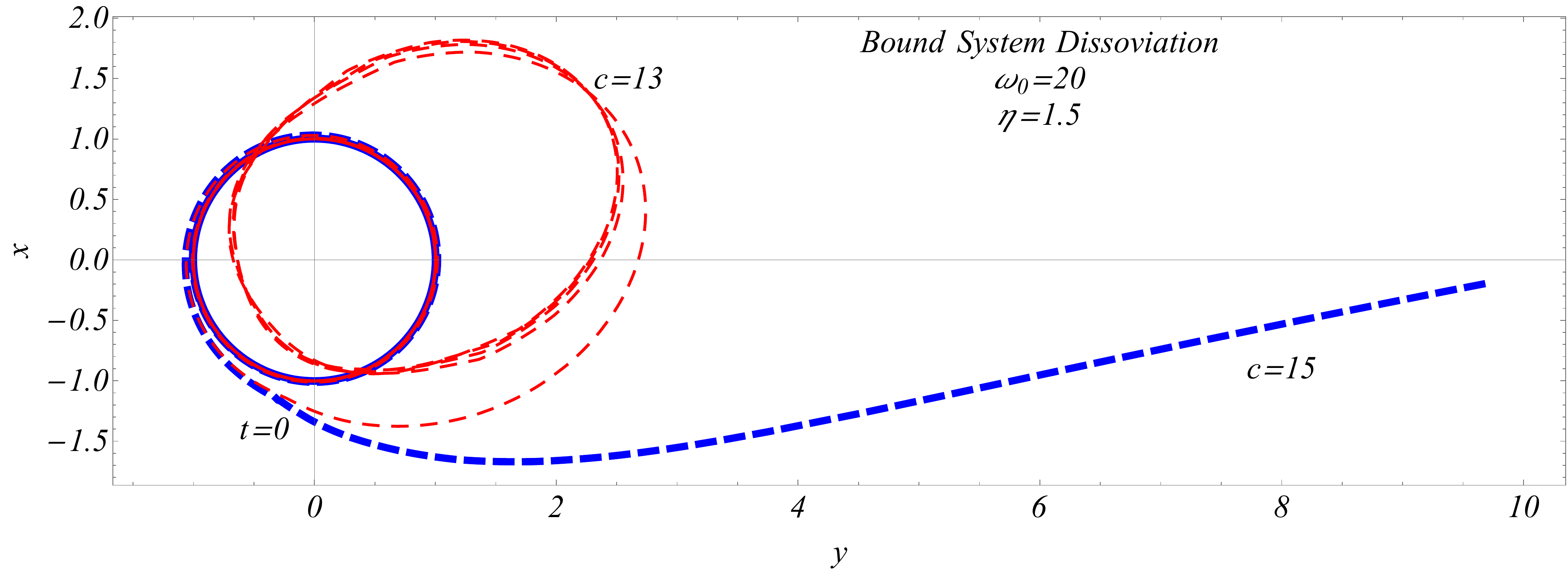}}}
\caption{The form of the geodesic bound system orbits for parameter $c$ values below (red line) and above (blue line) the critical value for dissociation which for $\omega_0=20$, $\eta=1.5$ is $c_{cr}\simeq 14$. The system clearly dissociates for $c>c_{cr}\simeq 14$ due to the outward impulse received at the time of the singularity ($t=0$). For $c>c_{cr}$ the system gets deformed from an initially circular to a final elliptical orbit (red line). The two lines initially overlap on the circular orbit. }
\label{fig6}
\end{figure}

The solutions $r(t)$ and $\varphi(t)$ are shown in Fig. \ref{fig5} for two values of the parameter $c$ with $\eta=1.5$ and $\omega_0=20$. The value $c=15$ (continous blue line) is above the critical value for dissociation $c_{cr}\simeq 14$ and the expansion impulse at the singularity provides enough energy to dissociate the system. This dissociation manifests itself as an unbounded increase of $r(t)$ for $t>0$ (Fig. \ref{fig5}a) while the angular coordinate remains constant (Fig. \ref{fig5}b). For $c=13$ (red dashed line) below the critical value for dissociation, the expansion impulse disrupts the bound system but it is not energetic enough to dissociate it. The radial coordinate becomes oscillatory while the angular coordinate continues to increase monotonically. The circular bound trajectory is disrupted to an elliptic one. This is shown more clearly in Fig. \ref{fig6} where we show the two trajectories in cartesian coordinates. At the time of the singularity, indicated on Fig. \ref{fig6} by the label `$t=0$', the initial circular orbit is transformed to either an elliptic orbit (for $c=13$) or to a free straight line trajectory (for $c=15$).

We now use energetic considerations to obtain an analytical estimate of the critical values $c_{cr}(\eta,\omega_0)$ of the parameter $c$ such that for $c>c_{cr}$ the expansion impulse due to the singularity is energetic enough to dissociate the bound system. Ignoring the contribution of the expansion away from the singularity, the binding energy of the system (depth of the effective potential (\ref{veff})) is
\be
V_{min}=-\frac{\omega_0^2}{2}
\label{vmin}
\ee 
The velocity change due to the expansion impulse is found by setting $r\simeq 1$ (approximate equilibrium radius) and integrating the expansion force in a large enough time interval $T$ around the singularity as
\ba 
\Delta {\dot r}(c,\eta,T)&\simeq & \int_{-T}^{T} \frac{\ddot a}{a}\; dt = 2 c\; \eta\; (\eta-1) \int_{0}^{T} \frac{t^{\eta-2}}{1+c\; t^\eta}\; dt= \nn \\&=&
2\; T^{\eta-1} c\; \eta\; _2F_1(1,1-\frac{1}{\eta},2-\frac{1}{\eta},-c\; T^\eta)
\label{velpert}
\ea
By demanding that the kinetic energy gained due to the expansion impulse is equal to the binding energy of the system we obtain 
\be 
\Delta {\dot r}(c,\eta,T)^2=\omega_0^2
\label{dissoccond}
\ee
where $T$ is assumed to be large enough to fully include the singularity.
\begin{figure}[!t]
\centering
\vspace{0cm}\rotatebox{0}{\vspace{0cm}\hspace{0cm}\resizebox{0.49\textwidth}{!}{\includegraphics{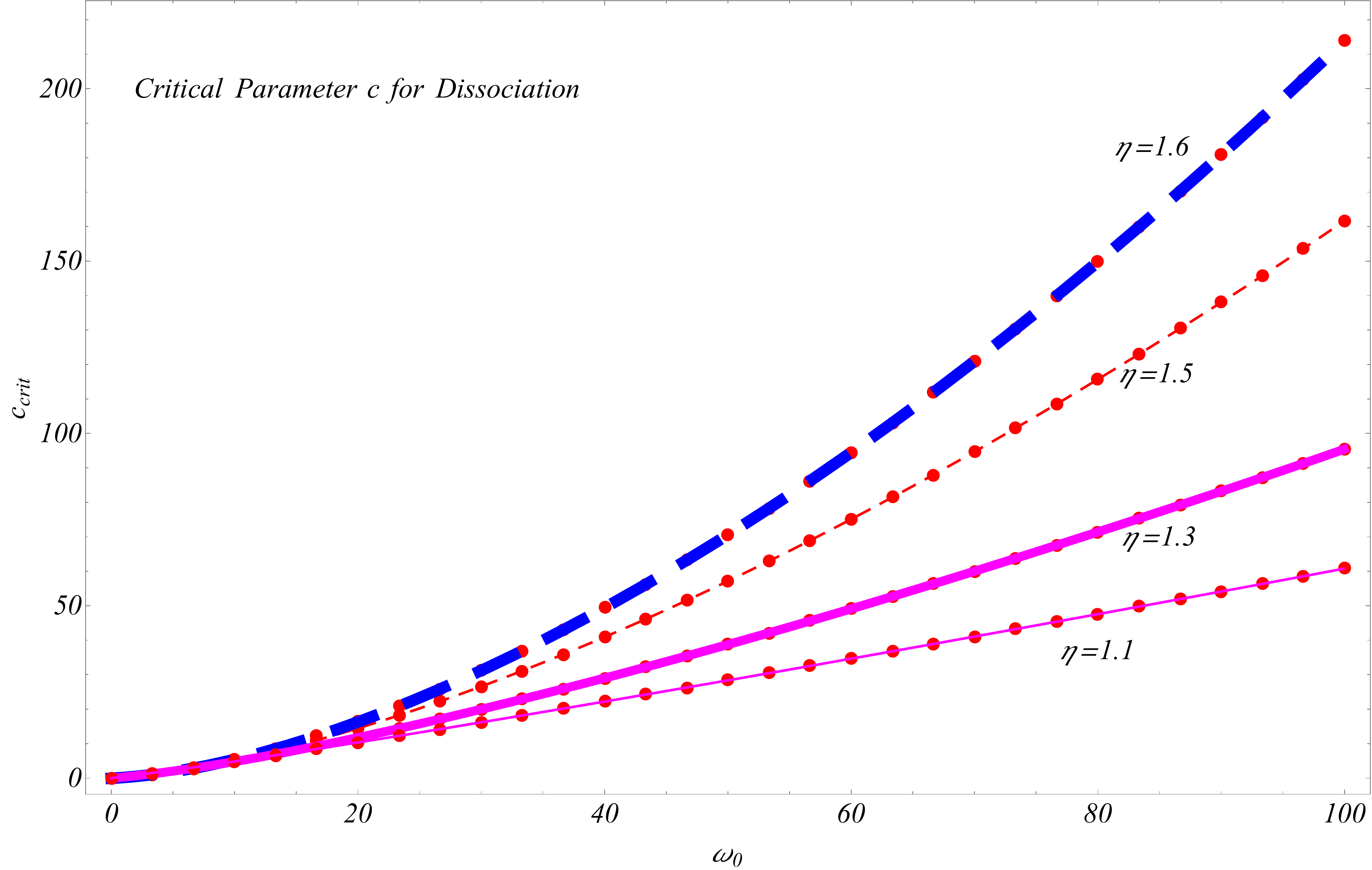}}}
\caption{The critical parameter $c_{cr}$ for bound system dissociation. The points were obtained using both the empirical formula and a numerical simulation confirmation while the lines interpolate through these points. Bound systems dissociate for $c>c_{cr}$. As expected more tightly bound systems (larger $\omega_0$) require a larger impulse to dissociate (larger value of $c$).}
\label{fig7}
\end{figure}

The solution of eq. (\ref{dissoccond}) can lead to a rough estimate of critical values $c_{cr}(\eta,\omega_0)$ required for bound system dissociation. Despite of the approximations involved in deriving eq. (\ref{dissoccond}) (eg ignoring the expansion effects in the binding energy and assuming fixed radius) we have found that the values of $c_{cr}(\eta,\omega_0)$ obtained by numerical solution of the geodesic equation (\ref{sfsgeod}) differ by only about $10-30\%$ from the estimate obtained using the analytical arguments of eq. (\ref{dissoccond}).

The values of  $c_{cr}(\eta,\omega_0)$ obtained by numerical solution of the geodesic equation (\ref{sfsgeod}) for various values of $c$, are shown in Fig. \ref{fig7}. For the construction of Fig. \ref{fig7} we obtained numerically the geodesic trajectories similar to those shown in Figs \ref{fig5}, \ref{fig6} in order to determine the critical values $c_{cr}$ such that for $c>c_{cr}$ the orbit is transformed at the singularity from bound to unbound. 

The construction of Fig. \ref{fig7} was aided by using empirical relations which emerge as modifications of eq. (\ref{dissoccond}) and provide a more accurate determination of $c_{cr}(\eta,\omega_0)$. For example a fairly accurate such empirical relation is of the form
\be 
2\Delta {\dot r}(c_{cr},\eta,\lambda_1(\eta) c^{1-/\eta})^2=\omega_0^2
\label{empir1}
\ee
where $0<\lambda_1(\eta)<1$ is obtained by demanding agreement with the numerical results and is found to depend weakly only on $\eta$ while it is independent of $\omega_0$. For example we have found that $\lambda_1(1.5)\simeq \lambda_1(1.6)=0.85$ while $\lambda(1.1)=0.075$. For such values of $\lambda_1$ the roots of eq. (\ref{empir1}) lead to the correct values of $c_{cr}$ shown in Fig. \ref{fig7} within about $3\%$ for all values of $\omega_0$ shown in Fig. \ref{fig7}.

As shown in Fig. \ref{fig7}, as the value of 
$\eta$ decreases towards $\eta=1$ the value of $c_{cr}$ decreases towards $0$ and for $\eta=1$ we have $c_{cr}=0$ implying that for $\eta\leq 1$ all values of $c>0$ lead to bound system dissociation as expected due to the diverging impulse induced by the expansion for this range of $\eta$.

\begin{figure}[!t]
\centering
\vspace{0cm}\rotatebox{0}{\vspace{0cm}\hspace{0cm}\resizebox{0.49\textwidth}{!}{\includegraphics{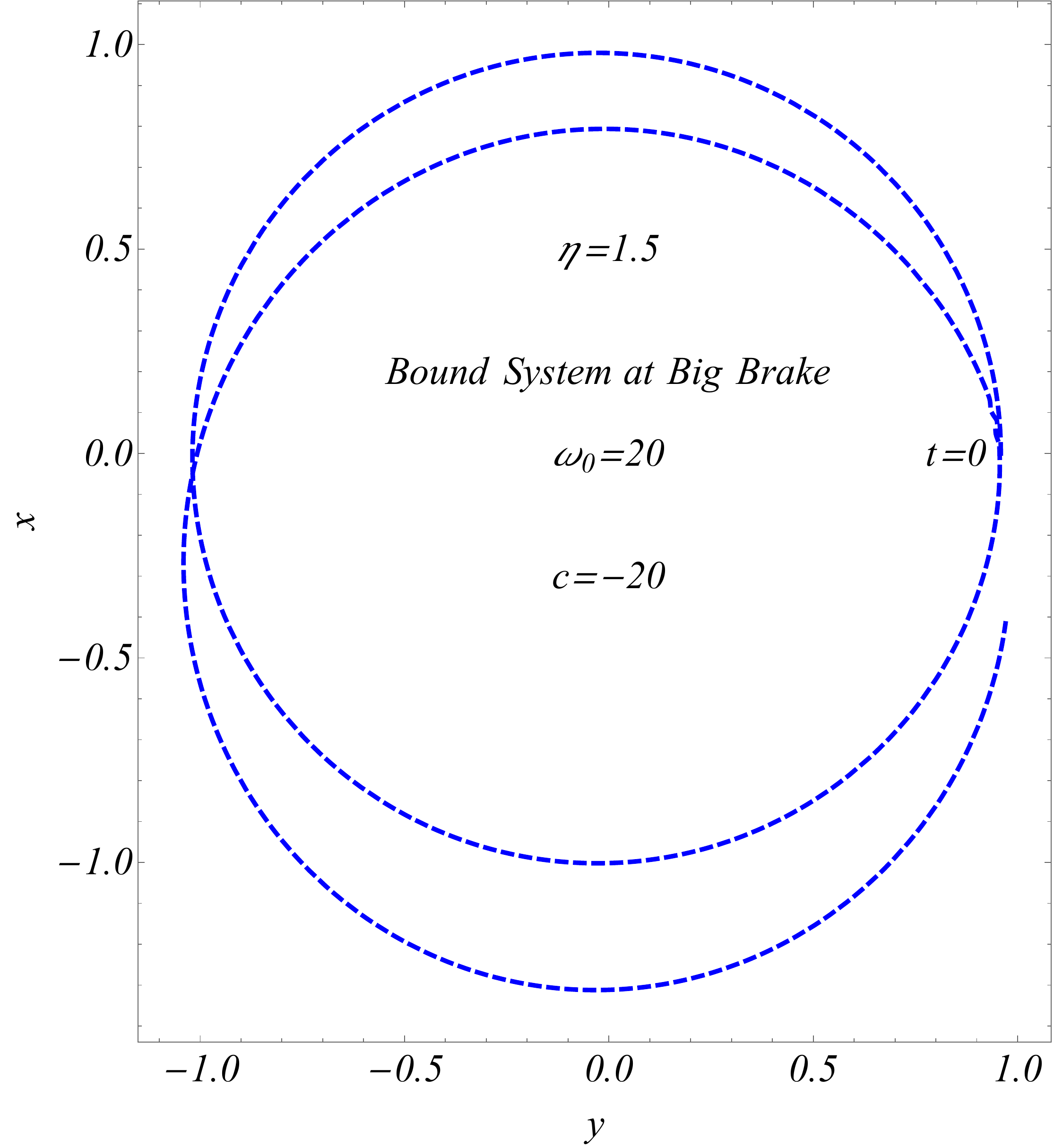}}}
\caption{For values of $c<0$ (Big Brake singularity) the impulse received by the bound system at $t=0$ is towards the center of the circular orbit and dissociation does not seem to occur for any value of $c$. Instead we have disruption of the circular orbit.}
\label{fig8}
\end{figure}

For $c<0$ the impulse of the expansion force is towards the center of the circular orbit leading to deformation of the system and no dissociation is observed for any value of $c$. The corresponding singularity for $1<\eta<2$ is known as Big Brake \citep{Keresztes:2010fi-tacyonic-bigbrake-tobogcrunch-obs-constr}. A typical deformation of the circular orbit through the Big Brake singularity is shown in Fig. \ref{fig8}. In the case of Big Brake the scale factor (\ref{scfactans}) has two roots (one before and one after the singularity) at
\be 
t_*=\pm \left( \frac{1}{\vert c \vert}\right)^{1/\eta}
\label{tstar}
\ee
corresponding to geodesically incomplete singularities (Big Bang and Big Crunch). The trajectory shown in fig. \ref{fig8} corresponds to the time range $t\in [-0.95 \vert t_* \vert, 0.95 \vert t_* \vert]$ and the Big Brake singularity occurs at the point close to the label $t=0$ where the discontinuity of the velocity is evident. For $\eta<1$ the discontinuity of the velocity diverges and the numerical construction of the orbit beyond the singularity was not possible.

\section{Conclusion}
\label{sec:Section 4}

We have derived analytically and numerically the form of free particle geodesics through SFS and demonstrated their existence when the scale factor is finite through the singularity.

We have also demonstrated that bound systems can survive through SFS provided that the impulse they receive at the singularity is less than a critical value which corresponds to a critical value of the parameter determining the form of the scale factor through the SFS. This critical parameter $c_{cr}(\eta,\omega_0)$ depends both on the exponent $\eta$ of the scale factor and on the mass and scale of the bound system through the parameter $\omega_0$. 

Bound systems that have survived through a SFS suffer deformations that may be detectable through cosmological observations. For example spiral galaxies that have gone through a SFS would have elongated and deformed spiral arms. 

The present analysis focuses on geodesically complete singularities which assume finite scale factor as is the case for SFS. Geodesically incomplete singularities where the scale factor is not finite (eg Big Rip) always lead to dissociation of all bound systems and have been studied in detail previously \cite{Nesseris:2004uj-fate-bound-systems-big-rip}. The fate of bound systems and the precise form of their geodesics, in other types of geodesically incomplete singularities (eg a Big Crunch) would be an interesting extension of this project.

The detailed form of the predicted deformation of many particle multi-orbit systems is an interesting extension of the present analysis. In the context of such an analysis and after comparison with the observed forms of bound systems like clusters and galaxies it may be possible to obtain bounds on the strength of possible SFS in our past light cone or to detect signatures of such events in the form of existing bound systems.

Another extension of the present analysis could be the investigation of the effects of SFS on cosmic defects like cosmic strings and domain walls in both the Nambu-Goto action approximation\cite{Balcerzak:2006ac} and in the full field theoretical formulation. Similar issues may be addressed regarding strongly bound systems like black holes \cite{Babichev:2004yx} where the approximate weak field metric we used is not applicable.

{\bf Numerical Analysis:} The Mathematica file that led to the production of the figures may be downloaded from \href{https://drive.google.com/open?id=0B7rg6X3QljQXYW1SVjgtejExNVU}{here}.

\section*{Acknowledgements}
I thank David Polarski and Antonio De Felice for fruitful discussions that led to the idea of this project.

\raggedleft
\bibliography{bibliography}

\end{document}